

\font\bigbold=cmbx12
\font\ninerm=cmr9      \font\eightrm=cmr8    \font\sixrm=cmr6
\font\fiverm=cmr5
\font\ninebf=cmbx9     \font\eightbf=cmbx8   \font\sixbf=cmbx6
\font\fivebf=cmbx5
\font\ninei=cmmi9      \skewchar\ninei='177  \font\eighti=cmmi8
\skewchar\eighti='177  \font\sixi=cmmi6      \skewchar\sixi='177
\font\fivei=cmmi5
\font\ninesy=cmsy9     \skewchar\ninesy='60  \font\eightsy=cmsy8
\skewchar\eightsy='60  \font\sixsy=cmsy6     \skewchar\sixsy='60
\font\fivesy=cmsy5     \font\nineit=cmti9    \font\eightit=cmti8
\font\ninesl=cmsl9     \font\eightsl=cmsl8
\font\ninett=cmtt9     \font\eighttt=cmtt8
\font\tenfrak=eufm10   \font\ninefrak=eufm9  \font\eightfrak=eufm8
\font\sevenfrak=eufm7  \font\fivefrak=eufm5
\font\tenbb=msbm10     \font\ninebb=msbm9    \font\eightbb=msbm8
\font\sevenbb=msbm7    \font\fivebb=msbm5
\font\tenssf=cmss10    \font\ninessf=cmss9   \font\eightssf=cmss8
\font\tensmc=cmcsc10

\newfam\bbfam   \textfont\bbfam=\tenbb \scriptfont\bbfam=\sevenbb
\scriptscriptfont\bbfam=\fivebb  \def\Bbb{\fam\bbfam}
\newfam\frakfam  \textfont\frakfam=\tenfrak \scriptfont\frakfam=%
\sevenfrak \scriptscriptfont\frakfam=\fivefrak  \def\frak{\fam\frakfam}
\newfam\ssffam  \textfont\ssffam=\tenssf \scriptfont\ssffam=\ninessf
\scriptscriptfont\ssffam=\eightssf  
\def\smc{\tensmc}

\def\eightpoint{\textfont0=\eightrm \scriptfont0=\sixrm
\scriptscriptfont0=\fiverm  \def\rm{\fam0\eightrm}%
\textfont1=\eighti \scriptfont1=\sixi \scriptscriptfont1=\fivei
\def\oldstyle{\fam1\eighti}\textfont2=\eightsy
\scriptfont2=\sixsy \scriptscriptfont2=\fivesy
\textfont\itfam=\eightit         \def\it{\fam\itfam\eightit}%
\textfont\slfam=\eightsl         \def\sl{\fam\slfam\eightsl}%
\textfont\ttfam=\eighttt         \def\tt{\fam\ttfam\eighttt}%
\textfont\frakfam=\eightfrak     \def\frak{\fam\frakfam\eightfrak}%
\textfont\bbfam=\eightbb         \def\Bbb{\fam\bbfam\eightbb}%
\textfont\bffam=\eightbf         \scriptfont\bffam=\sixbf
\scriptscriptfont\bffam=\fivebf  \def\bf{\fam\bffam\eightbf}%
\abovedisplayskip=9pt plus 2pt minus 6pt   \belowdisplayskip=%
\abovedisplayskip  \abovedisplayshortskip=0pt plus 2pt
\belowdisplayshortskip=5pt plus2pt minus 3pt  \smallskipamount=%
2pt plus 1pt minus 1pt  \medskipamount=4pt plus 2pt minus 2pt
\bigskipamount=9pt plus4pt minus 4pt  \setbox\strutbox=%
\hbox{\vrule height 7pt depth 2pt width 0pt}%
\normalbaselineskip=9pt \normalbaselines \rm}

\def\ninepoint{\textfont0=\ninerm \scriptfont0=\sixrm
\scriptscriptfont0=\fiverm  \def\rm{\fam0\ninerm}\textfont1=\ninei
\scriptfont1=\sixi \scriptscriptfont1=\fivei \def\oldstyle%
{\fam1\ninei}\textfont2=\ninesy \scriptfont2=\sixsy
\scriptscriptfont2=\fivesy
\textfont\itfam=\nineit          \def\it{\fam\itfam\nineit}%
\textfont\slfam=\ninesl          \def\sl{\fam\slfam\ninesl}%
\textfont\ttfam=\ninett          \def\tt{\fam\ttfam\ninett}%
\textfont\frakfam=\ninefrak      \def\frak{\fam\frakfam\ninefrak}%
\textfont\bbfam=\ninebb          \def\Bbb{\fam\bbfam\ninebb}%
\textfont\bffam=\ninebf          \scriptfont\bffam=\sixbf
\scriptscriptfont\bffam=\fivebf  \def\bf{\fam\bffam\ninebf}%
\abovedisplayskip=10pt plus 2pt minus 6pt \belowdisplayskip=%
\abovedisplayskip  \abovedisplayshortskip=0pt plus 2pt
\belowdisplayshortskip=5pt plus2pt minus 3pt  \smallskipamount=%
2pt plus 1pt minus 1pt  \medskipamount=4pt plus 2pt minus 2pt
\bigskipamount=10pt plus4pt minus 4pt  \setbox\strutbox=%
\hbox{\vrule height 7pt depth 2pt width 0pt}%
\normalbaselineskip=10pt \normalbaselines \rm}

\global\newcount\secno \global\secno=0 \global\newcount\meqno
\global\meqno=1 \global\newcount\appno \global\appno=0
\newwrite\eqmac \def\romappno{\ifcase\appno\or A\or B\or C\or D\or
E\or F\or G\or H\or I\or J\or K\or L\or M\or N\or O\or P\or Q\or
R\or S\or T\or U\or V\or W\or X\or Y\or Z\fi}
\def\eqn#1{ \ifnum\secno>0 \eqno(\the\secno.\the\meqno)
\xdef#1{\the\secno.\the\meqno} \else\ifnum\appno>0
\eqno({\rm\romappno}.\the\meqno)\xdef#1{{\rm\romappno}.\the\meqno}
\else \eqno(\the\meqno)\xdef#1{\the\meqno} \fi \fi
\global\advance\meqno by1 }

\global\newcount\refno \global\refno=1 \newwrite\reffile
\newwrite\refmac \newlinechar=`\^^J \def\ref#1#2%
{\the\refno\nref#1{#2}} \def\nref#1#2{\xdef#1{\the\refno}
\ifnum\refno=1\immediate\openout\reffile=refs.tmp\fi
\immediate\write\reffile{\noexpand\item{[\noexpand#1]\ }#2\noexpand%
\nobreak.} \immediate\write\refmac{\def\noexpand#1{\the\refno}}
\global\advance\refno by1} \def\semi{;\hfil\noexpand\break ^^J}
\def\nl{\hfil\noexpand\break ^^J} \def\refn#1#2{\nref#1{#2}}

\def\ann#1#2#3{{\it Ann.\ Phys.}\ {\bf {#1}} ({#2}) #3}

\def\jmp#1#2#3{{\it J.\ Math.\ Phys.}\ {\bf {#1}} ({#2}) #3}

\def\ijmp#1#2#3{{\it Int.\ J.\ Mod.\ Phys.}\ {\bf A{#1}} ({#2}) #3}

\def\pl#1#2#3{{\it Phys.\ Lett.}\ {\bf {#1}B} ({#2}) #3}
\def\plA#1#2#3{{\it Phys.\ Lett.}\ {\bf {#1}A} ({#2}) #3}

\def\prD#1#2#3{{\it Phys.\ Rev.}\ {\bf D{#1}} ({#2}) #3}
\def\prl#1#2#3{{\it Phys.\ Rev.\ Lett.}\ {\bf #1} ({#2}) #3}

\newif\iftitlepage \titlepagetrue \newtoks\titlepagefoot
\titlepagefoot={\hfil} \newtoks\otherpagesfoot \otherpagesfoot=%
{\hfil\tenrm\folio\hfil} \footline={\iftitlepage\the\titlepagefoot%
\global\titlepagefalse \else\the\otherpagesfoot\fi}

\def\abstract#1{{\parindent=30pt\narrower\noindent\ninepoint\openup
2pt #1\par}}

\newcount\notenumber\notenumber=1 \def\note#1
{\unskip\footnote{$^{\the\notenumber}$} {\eightpoint\openup 1pt #1}
\global\advance\notenumber by 1}

\def\today{\ifcase\month\or January\or February\or March\or
April\or May\or June\or July\or August\or September\or October\or
November\or December\fi \space\number\day, \number\year}

\def\pagewidth#1{\hsize= #1}  \def\pageheight#1{\vsize= #1}
\def\hcorrection#1{\advance\hoffset by #1}
\def\vcorrection#1{\advance\voffset by #1}

\pageheight{23cm}
\pagewidth{15.7cm}
\hcorrection{-1mm}
\magnification= \magstep1
\parskip=5pt plus 1pt minus 1pt
\tolerance 8000
\def\bsk{\baselineskip= 14.5pt plus 1pt minus 1pt}
\bsk

\font\extra=cmss10 scaled \magstep0  \setbox1 = \hbox{{{\extra R}}}
\setbox2 = \hbox{{{\extra I}}}       \setbox3 = \hbox{{{\extra C}}}
\setbox4 = \hbox{{{\extra Z}}}       \setbox5 = \hbox{{{\extra N}}}





\def\frac#1#2{{#1\over#2}}

\def\pmb#1{\setbox0=\hbox{$#1$} \kern-.025em\copy0\kern-\wd0
    \kern.05em\copy0\kern-\wd0 \kern-.025em\raise.0433em\box0 }
\def\noblackboxes{\overfullrule 0pt}

\def\({\left(}
\def\){\right)}

\def\figone{wallfig1.eps}
\def\figtwo{wallfig2.eps}
\def\figthree{wallfig3.eps}
\def\figfour{wallfig4.eps}
\input epsf

\let\omitpictures=N

\noblackboxes


{

\refn\MT
{C. Manuel and R. Tarrach, \pl{328}{1994}{113}}

\refn\CST
{T. Cheon, T. Shigehara and K. Takayanagi, {\it J. Phys. Soc. Japan}
{\bf 69} (2000) 345}

\refn\CS
{T. Cheon and T. Shigehara, {\it Phys. Lett.} {\bf A243} (1988) 111}

\refn\SMMC
{T. Shigehara, H. Mizoguchi, T. Mishima and T. Cheon,
{\it IEICE Trans. Fund. Elec. Comm. Comp. Sci.} {\bf E82-A} (1999) 1708}

\refn\RJ
{R. Jackiw, Paper I.3 in \lq\lq Diverse Topics in Theoretical and
Mathematical Physics\rq\rq, World Scientific, Singapore, 1995}

\refn\RT
{R. Tarrach, UB-ECM-PF 38-94, hep-th/9502020}

\refn\CSb
{T. Cheon and T. Shigehara, \prl{82}{1999}{2536}}

\refn\JPSJ
{I. Tsutsui, T. F\"ul\"op and T. Cheon, {\it J.\ Phys.\ Soc.\
Japan}\ {\bf 69} (2000) 3473--3476}

\refn\Ann
{T. Cheon, T. F\"ul\"op and I. Tsutsui, Symmetry, Duality and Anholonomy
of Point Interactions in One Dimension, \ann{294}{2001}{1}}

\refn\CHa
{T. Cheon, \plA{248}{1998}{285}}

\refn\JMP
{I. Tsutsui, T. F\"ul\"op and T. Cheon, M\"obius Structure of the Spectral
Space of Schr{\"o}dinger Operators with Point Interaction,
\jmp{42}{2001}{5687}}

\refn\AGHH
{S. Albeverio, F. Gesztesy, R. H{\o}egh-Krohn and H. Holden, \lq\lq
Solvable Models in Quantum Mechanics\rq\rq, Springer, New York, 1988}

\refn\Shi
{F. Shimizu, \prl{86}{2001}{987}}

\refn\FT
{T. F\"{u}l\"{o}p and I. Tsutsui, \plA{264}{2000}{366}}

\refn\FG
{E. Farhi and S. Gutmann, \ijmp{5}{1990}{3029}}

\refn\Gut
{M.C. Gutzwiller, ``Chaos in Classical and Quantum Mechanics'',
 Springer-Verlag, Heidelberg, 1991}

\refn\CMS
{T.E. Clark, R. Menikoff and D.H. Sharp, \prD{22}{1980}{3012}.
The possible bound state is not treated in this paper}

\refn\GP
{A. Galindo and P. Pascual, \lq\lq Quantum mechanics I\rq\rq,
Springer-Verlag, Heidelberg, 1990; p.154}

\refn\BO
{D. Boll\'e and T.A. Osborn, \prD{13}{1976}{299}}

\refn\LL
{L.D. Landau and E.M. Lifshitz, \lq\lq Mechanics\rq\rq, Course of
Theoretical Physics Vol.1, Butterworth-Heinemann,
Oxford, 1976}

\refn\Schulman
{L.S. Schulman, \lq\lq Techniques and Applications of
Path Integration\rq\rq, John Wiley \& Sons, New York, 1981}

}


\def\AABR{4.6}
\def\AAAB{3.2}



\def\m#1{$#1$}
\def\mm#1{$\,#1\,$}
\def\mmm#1{$\,\,#1\,\,$}
\def\mmmm#1{$\,\,\,#1\,\,\,$}
\def\sectit#1{\bigskip \noindent{\bf #1} \smallskip}
\def\mdef#1{ \left\{ \matrix{ #1 } \right. } 

\def\f#1#2{{#1\over#2}}
\def\ff#1#2{\raise.2pt\hbox{\ninepoint${\displaystyle\f{#1}{#2}}$}}
\def\fff#1#2{\raise.5pt\hbox{\eightpoint${\displaystyle\f{#1}{#2}}$}}
\def\ffff#1#2{{\textstyle \f{#1}{#2}}}
\def\F#1#2{#1/#2}
\def\FF#1#2{(#1/#2)}
\def\FFF#1#2{#1/(#2)}
\def\FFG#1#2{#1/[#2]}
\def\FGF#1#2{(#1)/#2}

\def\s#1{\sqrt{#1}}
\def\scal#1#2{{\langle #1, #2 \rangle}}
\def\quotes#1{`#1'}
\def\quotess#1{``#1''}

\def\pmbfh{0.035ex}  \def\pmbfv{0.05ex}
\def\pmbf#1{\hbox{$\setbox5=\hbox{$#1$}\copy5\kern-\wd5\raise\pmbfv\copy5
      \kern-\wd5\kern\pmbfh\kern\pmbfh\copy5\kern-\wd5\raise\pmbfv\box5$}}
\def\hpt#1{\hskip #1 truept}
\def\vpt#1{\vskip #1 truept}

\def\ph#1{\phantom{#1}}

\def\biTTTT{\hskip -0.60ex}
\def\biTTT{\hskip -0.45ex}
\def\biTT{\hskip -0.30ex}
\def\biT{\hskip -0.15ex}
\def\bi{\hskip 0.00ex}
\def\bit{\hskip 0.15ex}
\def\bitt{\hskip 0.30ex}
\def\bittt{\hskip 0.45ex}
\def\bitttt{\hskip 0.60ex}
\def\hf{\hfill}

\def\eg{{\it e.g.,\ }}
\def\ie{{\it i.e.,\ }}

\def\lhs{l.h.s.\ }
\def\rhs{r.h.s.\ }

\def\cf{cf.\ }
\def\text#1{{\rm #1}}
\def\textt#1{{\rm #1\,}}
\def\texttt#1{{\rm #1\ }}

\def\d{\text{d}}

\def\arccot{\textt{arccot}}
\def\const{\textt{const.}}

\def\mod{\texttt{mod}}

\def\phy{\varphi}
\def\eps{\varepsilon}

\def\p{\partial}
\def\h{\hbar}

\def\[{\left[}
\def\]{\right]}
\def\<{\langle}
\def\>{\rangle}
\def\=>{\Rightarrow}
\def\==>{\Longrightarrow}
\def\co{\, ,}

\def\pe{\, .}
\def\upto{\nearrow}
\def\downto{\searrow}

\def\case#1{\medskip\noindent\underbar{$\pmbf{\hbox{#1}}$:}}

\def\phsh{\delta}
\def\taucl{\tau_{\text{cl}, \bitt x_0}}
\def\xd{x_{\text{disc}}}
\def\xk{x_{\text{wall}}}
\def\xl{x_{\text{thresh}}}
\def\xo{x_{\text{pos}}}
\def\xinf{x_{\infty}}
\def\xone{x_1}
\def\xtwo{x_2}
\def\Etwo{E_2}
\def\Eo{E_*}
\def\qn{\eta(x)}

\def\qk{\tilde{k}}
\def\qc{c}
\def\qnu{\nu}

\def\Eb{E_{\rm bounce}}
\def\Ed{E_{\rm direct}}
\def\Sb{S_{\rm bounce}}
\def\Sd{S_{\rm direct}}

\def\++{^{(+)}}
\def\--{^{(-)}}

\null

{
\leftskip=100mm
\hfill\break
KEK Preprint 2001-134
\hfill\break
\null \hpt{29} quant-ph/0111057
\hfill\break
\par
}

\vskip 10pt

\centerline{\bigbold
Classical Aspects of Quantum Walls in One Dimension}
\vskip 30pt

\centerline{\smc
Tam\'{a}s F\"{u}l\"{o}p\footnote{${}^*$}
{\eightpoint email:\quad fulopt@poe.elte.hu}
}

\vskip 3pt
{
\baselineskip=13pt
\centerline{\it
Institute of Particle and Nuclear Studies}
\centerline{\it High Energy Accelerator Research Organization (KEK)}
\centerline{\it Tsukuba 305-0801,
Japan
}

\vskip 10pt

\centerline{\smc
Taksu Cheon\footnote{${}^\dagger$}
{\eightpoint email:\quad cheon@mech.kochi-tech.ac.jp,
http://www.mech.kochi-tech.ac.jp/cheon/}  }

\vskip 3pt
{
\baselineskip=13pt
\centerline{\it Laboratory of Physics}
\centerline{\it Kochi University of Technology}
\centerline{\it Tosa Yamada, Kochi 782-8502, Japan}
}

\vskip 7pt
\centerline{\rm and}
\vskip 3pt

\centerline{\smc Izumi Tsutsui\footnote{${}^\ddagger$}
{\eightpoint email:\quad izumi.tsutsui@kek.jp,
http://research.kek.jp/people/itsutsui/}  }

\vskip 3pt

{
\baselineskip=13pt
\centerline{\it
Institute of Particle and Nuclear Studies}
\centerline{\it
High Energy Accelerator Research Organization (KEK)}
\centerline{\it
Tsukuba 305-0801, Japan}
}
\vskip 45pt

\abstract{%
{\bf Abstract.} \quad
We investigate the system of a particle moving on a half line $x \ge 0$
under the general walls at $x = 0$ that are permitted quantum
mechanically.  These quantum walls, characterized by a parameter $L$, are
shown to be realized as a limit of regularized potentials.  We then study
the classical aspects of the quantum walls, by seeking a classical
counterpart which admits the same time delay in scattering with the
quantum wall, and also by examining the WKB-exactness of the transition
kernel based on the regularized potentials.  It is shown that no classical
counterpart exists for walls with $L < 0$, and that the WKB-exactness can
hold only for $L = 0$ and $L = \infty$.
}

\vskip 10pt
{\baselineskip=10pt
{\ninepoint
}
}

\vfill\eject


\pageheight{23cm}
\pagewidth{15.7cm}
\hcorrection{-1mm}
\magnification= \magstep1
\def\bsk{%
\baselineskip= 14pt plus 1pt minus 1pt}
\parskip=5pt plus 1pt minus 1pt
\tolerance 8000
\bsk


\sectit{1. Introduction}

\secno=1 \meqno=1

\medskip

Quantum systems with contact interactions ({\it i.e.}, point interactions
or reflecting boundaries) enjoy an increasing interest recently.  On the
theoretical side, they have been found to exhibit a number of intriguing
features, many of which have been seen before only in connection with
quantum field theories. Examples include renormalization [\MT, \CST, \CS,
\SMMC, \RJ], Landau poles [\RT], anomalous symmetry breaking [\RJ],
duality [\CSb, \JPSJ, \Ann], supersymmetry [\Ann] and spectral anholonomy
[\Ann, \CHa, \JMP].  On the experimental side, the rapid developments of
nanotechnology forecast that nano-scale quantum devices can be designed
and manufactured into desired specifications.  The description of some of
these systems will involve the theory of contact interactions.  As a
simple example, a piece of a single nanowire would act as a one
dimensional line with two reflecting endpoints between which a conduction
particle moves almost freely, allowing for a quantum mechanical
description with boundaries.  Other applications arise, for instance, in
systems with impurities which act as point scatterers.  All these areas of
interest lend impetus to investigate quantum systems with contact
interactions further to uncover their full potential both theoretically
and experimentally.

The topic of this paper is the quantum half line system, which is perhaps
the simplest among those with contact interactions.  This system also
appears frequently as the radial part of higher dimensional systems
[\AGHH]. (For the recent experimental studies, see [\Shi] and references
therein.) We consider a quantum particle that moves freely on a half line
$x \ge 0$ with the endpoint $x = 0$ acting as a reflecting boundary, or an
impenetrable wall. This system is known (see section 2) to admit a
one-parameter family of distinct walls characterized by the boundary
conditions,
    $$
    \psi(0) + L \, \psi'(0) = 0\ ,
    \eqn\aaag
    $$
where $L$ is a parameter which takes all real numbers including $L =
\infty$.  Clearly, the standard wall in which we impose $\psi(0) = 0$ is
obtained for $L = 0$ but it is just one of the various walls allowed, and
therefore the first question one may ask is whether those nonstandard
walls with $L \ne 0$ can arise in actual physical settings.

To answer this, we study how those nonstandard walls can be realized as a
limit of finite (regularizing)  potentials.  The potentials we consider
are step-like and may readily be manufactured using, \eg thin layers of
different types of semiconductors.  We shall show that it is indeed
possible to realize such nonstandard walls out of the step-like potentials
if we fine-tune the limiting procedure.  We then turn to the question
whether such nonstandard walls are available only quantum mechanically or
not.  This will be examined by looking at the time delay of the particle
in scattering, which is the time difference between the moments of
incidence and reflection at the wall.  It will be shown that quantum
nonstandard walls with $L < 0$, which are characterized by positive time
delay, have no classical counterpart possessing the same time delay, which
implies that these walls are purely quantum. We also consider the validity
of the semiclassical WKB approximation for the transition kernel under
nonstandard walls, where now one takes into account the possible two
classical paths, the direct path and the bounce path, in the path integral
[\FT].  This is of interest because it has been known that, for the
standard wall as well as that of $L = \infty$, the WKB approximation
becomes exact if a sign factor is properly attached to the contribution of
the bounce path.  We shall see that for these two values of $L$ the
required sign factor can be accounted for by the bounce effect, showing
that the WKB approximation is in fact exact, whereas for other $L$ the
WKB-exactness cannot hold. Before presenting these results, we provide the
basics of the quantum system on the half line below.

\sectit{2. Basics of the quantum system on the half line}

\secno=2 \meqno=1

The system of a (nonrelativistic) free particle on a half line $x \in [0,
\infty)$ is governed by the Hamiltonian \mm{ H = - \FFF{\h^2 \biTTT}{\biT
2m} \bitttt \F{\d^2 \biTT}{\d x^2} \co } supplemented by some boundary
condition imposed at the wall $x = 0$.  The boundary condition is
determined by the requirement that $H$ be self-adjoint on the positive
half line $x \ge 0$ and, mathematically, this is done by finding proper
domains of the operator $H$ on which it is self-adjoint.  The result is
that there exists a $U(1)$ family of domains of states specified by
(\aaag)  (see, \eg [\AGHH], Appendix D), which can be readily understood
by a direct inspection as well.  Indeed, one sees by partial integration
that for $H$ to be self-adjoint one must have \mmm{\psi^* \psi' = \psi'^*
\psi} at \mm{x = 0} for any state $\psi$ on which $H$ acts.  If $\psi'(0)
\ne 0$, this implies $ \; \psi(0) / \psi'(0) = [ \psi(0) / \psi'(0) ]_{}^*
= - L \; $ with $L$ being some real constant, which is just the condition
(\aaag).  \note{The fact that the constant $L$ is universal for any state
$\psi$ can be seen by considering (\aaag) for all linear combinations of
two states $\psi_1$ and $\psi_2$ with $L_1$ and $L_2$, from which one
deduces $L_1 = L_2$ immediately.} The case $\psi'(0) = 0$ which also
fulfills the requirement can be included by allowing $L = \infty$ in
(\aaag).  The whole family is $U(1)$ because of the range of the
parameter:  $L \in ( - \infty, \infty ) \cup \{ \infty \} \cong U(1) $.

Under the boundary condition (\aaag)  the positive energy states are
    $$
    \phy_k (x) = \fff{1}{ \sqrt{2\pi} } \( e^{-ikx} +
    e^{i \phsh_k} e^{ikx} \)
    \eqn\aaaq
    $$
with \mm{ \phsh^{}_k = 2 \,\arccot kL}.
In addition, for $L > 0$, we also have one negative energy state,
    $$
    \phy_{\rm bound} (x) =
    \sqrt{ \ffff{2}{L} } \, \, e^{ - \frac{x}{L} } \hpt{40} (L > 0) \co
    \eqn\aaax
    $$ which is a bound state localized at the wall with its
characteristic size $L$.  The existence of the bound state (\aaax) can
also be ensured from the minimum energy condition.  Namely, for any
normalized state $\psi$ the expectation value of the energy reads
     $$
     \scal{\psi}{H \psi} = \fff{\hbar^2}{2m} \fff{1}{ L^2}
     \int_0^\infty \d x \, | \psi(x) +  L \psi'(x) |^2 -
     \fff{\hbar^2}{2m} \fff{1}{ L^2}\ ,
     \eqn\aabc
     $$
where $L$ is the parameter in (\aaag). The lower bound \mm{ - \f{\h^2}{2m}
\f{1}{L^2} } is attained if there exists a state satisfying \mm{ \psi(x) +
L \psi'(x) = 0} for all $x \ge 0$, which is just the bound state (\aaax).

As seen in the bound state, the parameter $L$ furnishes a physical scale
in many of the properties of the system. An example for this is provided
by the time delay that occurs when an incoming particle is reflected from
the wall. The time delay in quantum scattering processes has been studied
extensively (see, \eg [\GP, \BO] and references therein). Its definition
and calculation is done for our system as follows.\note{Compare this with
the classical mechanical definition of time delay, presented in Sect.~4.}
Let us consider a wave packet formed out of the positive energy states
(\aaaq),
    $$
    \eqalign{ \psi(x,t) &= \int_0^\infty \d k \,f(k) \,
    e^{i k x_0} e^{- \f{i \h k^2}{2m} t } \phy_k(x) \cr
    & = \fff{1}{\s{2\pi}} \int_0^\infty \d k \, f(k) \,
    e^{i k x_0} e^{- \f{i \h k^2}{2m} t } e^{-ikx}  +
    \fff{1}{\s{2\pi}} \int_0^\infty \d k \, f(k) \, e^{i k x_0}
    e^{i \phsh_k} e^{- \f{i \h k^2}{2m} t } e^{ikx} }
    \eqn\packet
    $$
where $f(k)$ is a real function peaked at $k_0 > 0$.  The first term
describes the incident packet whose maximum starts from $x_0$ at $t = 0$
and moves to the left with velocity magnitude \mm{ v_0 = \F{\hbar k_0
\biT}{m} \co } as can be seen from a stationary phase argument,
    $$
    \left. \F{\d}{\d k} \left( - \FFF{\hbar k^2 \biTTT}{2m} \bittt
    t + k x_0 - k x \right)
    \right| \biTTT \raise-3.4pt\hbox{$ {}_{k = k_0} $}
    = 0 \hpt{18} \==> \hpt{18} x_{{\rm max}}^{(1)} (t) =
    x_0 - \FF{\hbar k_0 \biT}{m} \, t \pe
    \eqn\xmaxone
    $$
Similarly, the reflected packet given by the second term moves as
    $$
    x_{{\rm max}}^{(2)} (t) = - x_0 + \FF{\hbar k_0 \biT}{m} \bitt t +
    \FFG{2L}{ 1 + (k_0 L)^2 } \pe
    \eqn\xmaxtwo
    $$
As $t$ increases, the first packet moves towards the wall at $x = 0$, and
its maximum reaches it at $ \; t_1 = x_0 / v_0 \; $.  Meanwhile, the
second packet comes from the left (if we allow $x < 0$ as well)  moving
to the right and arrives at the wall at $ \; t_2 = ( x_0 - \frac{2L}{1 +
(k_0 L)^2 } ) / v_0 $.  The difference between the two instants gives the
time delay,
    $$
    \tau = t_2 - t_1 = - \frac{2mL}{ \hbar k_0 [ 1 + (k_0 L)^2 ] } \pe
    \eqn\td
    $$
For $L = 0$ and $L = \infty$, this time delay is zero, as one would expect
on the ground that for such cases there is no parameter in the system
possessing the dimension of time.  Note that for negative $L$ the time
delay is positive, whereas for positive $L$ it is negative.

{}From the eigenfunctions (\aaaq) and (\aaax) the Feynman kernel
describing the transition of the particle from $x = a$ at $t = 0$ to $x =
b$ at $t = T$ can be calculated (see [\FG, \Gut, \CMS]).  The result is
    $$
    K(b, T; a, 0) = \s{ \fff{m}{2 \pi i \h T} } \[ e^{ \f{i m}{2 \h T}
    (b - a)^2 } \mp e^{ \f{i m}{2 \h T} (b + a)^2 } \]\ ,
    \eqn\aabd
    $$
{}for $L = 0$ (\quotes{$-$}-sign) and $L = \infty$
(\quotes{$+$}-sign).  For $L < 0$ the kernel is given by
    $$
    \s{ \fff{m}{2 \pi i \h T} } \bittt \bigg[ e^{ \f{i m}{2 \h T}
    (b - a)^2 } + e^{ \f{i m}{2 \h T} (b + a)^2 } - \fff{2}{|L|}
    \int_0^\infty \d z \, e^{- z / |L|} \,
    e^{ \f{i m}{2 \h T} (b + a + z)^2 } \bigg]\ ,
    \eqn\aabe
    $$
and for $L > 0$ by
    $$
    \s{ \fff{m}{2 \pi i \h T} } \bittt \bigg[ e^{ \f{i m}{2 \h T}
    (b - a)^2 } + e^{ \f{i m}{2 \h T} (b + a)^2 } - \fff{2}{L}
    \int_0^\infty \! \d z \, e^{- {z}/{L} } \, e^{ \f{i m}{2 \h T}
    (b + a - z)^2 } \bigg] + \fff{2}{L}
    e^{ \f{ i \h T }{ 2 m L^2 } } e^{ - \f{b + a}{L} } \pe
    \eqn\aabf
    $$
The salient feature of the result is that, for $L = 0$ and $L = \infty$,
the kernel (\aabd) almost coincides with that obtained by WKB
semiclassical approximation, because the two terms in (\aabd) correspond
to the free kernels for the direct path from $(a, 0)$ to $(b,T)$ and for
the bounce path which hits the wall once during the transition,
respectively.  The only problem for the complete WKB-exactness is the
appearance of the $\mp$ sign factor attached to the contribution from the
bounce path.  We shall show later that this sign factor can be attributed
to the classical action \mm{ \Delta \Sb = \hbar \pi} gained by the bounce
effect at the wall so that \mm{ e^{\f{i}{\hbar} \Delta \Sb} = \mp 1}.

\sectit{3. Realization of the wall}

\secno=3 \meqno=1

We now discuss how to realize the wall characterized by (\aaag) in actual
physical settings.  {}For this, we shall adopt a regularization method
which is analogous to those used earlier for point singularities [\SMMC,
\AGHH].  We extend the space to the entire line $-\infty < x < \infty$ and
seek a potential $V(x)$ with finite support such that, in the limit of
vanishing support, the boundary condition (\aaag) at $x = 0$ can be
realized.  Obviously, since no probability flow is admitted through the
wall at $x = 0$, such a regularized potential has to become infinitely
high for $x < 0$ in the limit.  A simple choice for the potential
fulfilling the demand is
    $$
    V(x) = \left\{ \matrix{ V_1 \co  & \ph{.} &  \ph{x} x < - d  &
    \ph{-} &
    \hbox{(domain I)} \hf  \cr
    V_2 \co  & &  - d < x < 0  & &
    \hbox{(domain II)} \hf  \cr
    0 \co  & &  \hskip 2pt x > 0  & &  \hbox{(domain III)} \hf  } \right.
    \eqn\aaaa
    $$
with constants $V_1 > 0$ and $V_2 < 0$.  Here, the scale of the support is
given by the regularization parameter $d$, and $V_1$ and $V_2$ are assumed
to be functions of $d$ such that $V_1, \, |V_2| \to \infty$ as $d \to 0$.

\topinsert
\epsfxsize 4.3cm
\ifx\omitpictures N  \centerline{\epsfbox {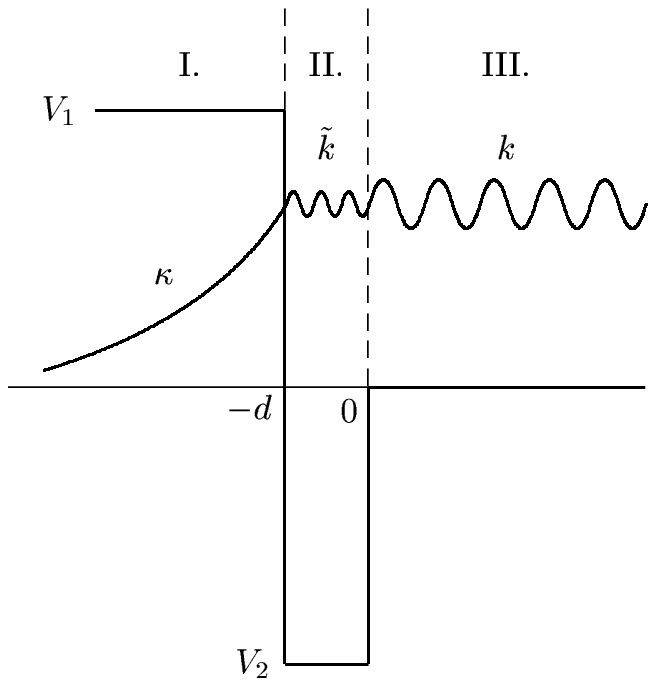}}  \fi
\vpt{5}
\centerline{\abstract{{\bf Figure 1.}~The regularized potential
(\aaaa) and the eigenfunction (\AAAB).}}
\bigskip
\endinsert

To find the appropriate dependence of $V_1(d)$ and $V_2(d)$, let us
consider an energy eigenstate $\phy$ in the potential (\aaaa) with energy
$E < V_1$ (see Figure~1):
    $$
    \phy(x) = \left\{ \matrix{
    \phy^{}_{\rm I}(x) = N e^{\kappa x}, \hskip 9ex & \ph{x} x < - d\ , &
    \quad  \, \kappa = \sqrt{ \f{2m}{\hbar^2} (V_1 - E) } \,
    \co \hf \cr
    \phy^{}_{\rm II}(x) = A e^{i \qk x} + B e^{-i \qk x}, &
    - d < x < 0\ , & \quad \, \qk = \sqrt{ \f{2m}{\hbar^2} (|V_2| + E) }
    \, \co \hf \cr
    \phy^{}_{\rm III}(x) = C e^{i k x} + D e^{-i k x}, &  x > 0 \ , &
    \quad \, k = \sqrt{ \f{2mE}{\hbar^2} } } \right.
    \eqn\aaab
    $$
\big(for \mm{E < 0}, \mm{ \phy^{}_{\rm III}(x) = M e^{- \sqrt{
\f{2m|E|}{\h^2} } x} }\big).  Under such finite potentials ({\it i.e.},
without infinity or singularity), the wave function and its derivative are
required to be continuous.  The condition which is dynamically important
is provided by the continuity of the ratio $\phy'/\phy$ which is free from
the ambiguity of overall normalization.  {}From this continuity condition,
we obtain
    $$
    \kappa = \frac{ i \qk ( A e^{-i \qk d} -
    B e^{i \qk d} ) }{ A e^{-i \qk d} +
    B e^{i \qk d} } \co \hskip 50pt
    \frac{ \phy_{\rm III}' }{ \phy_{\rm III} } \bitt (0) =
    \frac{ i \qk (A - B) }{ A + B }
    \eqn\aaad
    $$
at $ x = - d \bit $ and $x = 0$. Note that both $\qk$ and $\kappa$ are
$d$-dependent $\qk = \qk(d)$, $\kappa = \kappa(d)$ through $V_1(d)$ and
$V_2(d)$ and so are the two ratios in (\aaad). If we introduce
    $$
    R(d) = {{\phy_{\rm III}'}\over{\phy_{\rm III}}} \bitt (0) \ , \qquad
    \alpha = \arctan \f{\kappa}{\qk}\ , \qquad \beta = \qk d\ ,
    \eqn\radfn
    $$
then from (\aaad) we find
    $$
    R(d) =
    \qk \bitt \f{ ( A e^{-i \beta} - B e^{i \beta} ) \cos \beta
    - i ( A e^{-i \beta} + B e^{i \beta} ) \sin \beta }{
    ( A e^{-i \beta} + B e^{i \beta} ) \cos \beta -
    i ( A e^{-i \beta} - B e^{i \beta} ) \sin \beta }
    = \qk \tan (\alpha - \beta) \pe
    \eqn\aaae
    $$
The boundary condition (\aaag) is realized if
    $$
    R(d) \to -\fff{1}{L} \qquad \hbox{as} \quad d \to 0\ ,
    \eqn\aadl
    $$
independently of the energy $E$. In what follows we present a set of
regularized potentials fulfilling this requirement.

To this end, we first define
    $$
    \alpha_0 = \lim_{d \to 0} \alpha\ , \qquad
    \beta_0  = \lim_{d \to 0} \beta\ ,
    \eqn\ablmt
    $$
and note that, since \mm{V_1(d) \to \infty} as $d \to 0$, we always have
\mm{\kappa \to \infty}, whereas since \mm{0 < \alpha < {\pi}/{2}} by
definition, we have $0 \le \alpha_0 \le {\pi}/{2}$.  Note also that, if
$V_2(d)$ used in our regularization is such that \mm{\beta \to \infty},
then \mm{ \tan (\alpha - \beta) } will oscillate between \mm{-\infty} and
\mm{\infty} so \mm{R(d)} will not have a limit.  We therefore confine
ourselves to cases in which \mm{\beta} has a finite (zero or nonzero)
limit $\beta_0$.  Now, let us suppose $\beta_0 \ne \alpha_0 \bitttt (\mod
\pi)$, that is, $\tan( \alpha - \beta) \to \tan( \alpha_0 - \beta_0 ) \ne
0$.  Then, if \mm{\vert V_2\vert \to \infty} we have \mm{\qk \to \infty}
and, consequently, \mm{R(d) \to \pm \infty}.  If \mm{\vert V_2\vert }
remains finite, on the other hand, then we find \mm{\alpha_0 = \pi/2} and
\mm{\beta_0 = 0} and hence \mm{R(d)  \to \infty}.  We thus see that these
regularizations yield necessarily the standard wall \mm{L = 0}.

The foregoing argument shows that nonstandard walls with \mm{L
\ne 0} can be realized only by such realizations in which \mm{V_1} and
\mm{V_2} are fine-tuned as
    $$
    \beta_0 = \alpha_0 \bitttt (\mod \pi) \pe
    \eqn\aadn
    $$
We shall suppose (\aadn) from now on, and consider the limit of \mm{R(d)}
for the cases \mm{\alpha_0 = 0}, \mm{ 0 < \alpha_0 < \pi/2 } and
\mm{\alpha_0 = \pi/2}, separately.

\case{(i) case \mm{\alpha_0 = 0}}

We then have, as $d \to 0$, \mm{\alpha \approx \tan \alpha = \kappa / \qk
\to 0} and \mmm{\beta - \beta_0 \to 0} and hence \mm{ \tan ( \alpha -
\beta ) = \tan ( \alpha - \beta + \beta_0 ) \approx \F{\kappa}{\qk} -
\beta + \beta_0 \pe } Thus the ratio is approximated as
    $$
    R(d) \approx \kappa - \qk (\beta - \beta_0) \pe
    \eqn\aadp
    $$
Now, if \mm{\beta_0 = 0} then the \rhs reads \mm{\kappa - \qk^2 d}. Hence,
to get a finite \mm{R(d)}, \mm{\qk^2 d} has to compensate the divergence
of \mm{\kappa}. This can be done if $\kappa$ and $\qk$ behave as
    $$
    \kappa \sim \qc \bitt d^{\bit \qnu} - \fff{1}{L} \co \hskip 30pt
    \qk \sim \qc^{ \f{1}{2} } \bitt d^{ \f{\qnu - 1}{2} }
    \hskip 30pt ( -1 < \qnu < 0 ) \co
    \eqn\aadq
    $$
which is realized if, for instance, we put
    $$
    V_1 (d) = \fff{\h^2}{2m} \( \qc^2 d^{\bit 2 \qnu} - \fff{2 \qc}{L}
    \bit d^{\bit \qnu} \) \co \hskip 25pt V_2 (d) = - \fff{\h^2}{2m}
    \bitt \qc \bittt d^{\bit \qnu - 1} \co
    \eqn\aadr
    $$
with a constant $c > 0$.  It is then readily confirmed that this
regularized potential (\aadr)  does lead to $R(d)$ fulfilling (\aadl) for
all $E > 0$. If \mm{\beta_0 > 0}, on the other hand, then \mm{\beta_0
d^{-1} (\beta - \beta_0) } on the \rhs of (\aadp) has to cancel the
divergence of \mm{\kappa}. This means \mmm{ \qk \sim \beta_0 \bit d^{-1} +
(1 / \beta_0)  \bit \kappa \pe } The needed finite term \mm{ -\fff{1}{L} }
can be provided again by \mm{\kappa} if \mmm{ \kappa \sim c_1 d^{\qnu} -
\fff{1}{L}}. This is achieved, for example, by
    $$
    V_1(d) = \fff{\h^2}{2m} \( \qc^2 d^{2 \qnu} - \fff{2 \qc}{L} \bitt
    d^{\qnu} \) \co \hskip 25pt V_2(d) = - \fff{\h^2}{2m} \( \beta_0^2
    \bitt d^{-2} + 2 \qc \bitt d^{\qnu - 1} \) \pe
    \eqn\aadm
    $$
It is again easy to confirm that (\aadm) yields $R(d)$ fulfilling
(\aadl) for \mm{ \qnu > -1/2 }.

\case{(ii) case \mm{ 0 < \alpha_0 < \pi/2 }}

In this case, we have \mm{ \qk \sim \beta_0 d^{-1}} and \mm{ \kappa \sim
(\beta_0 \tan \beta_0) \bitt d^{-1} \pe } Using the Taylor expansion,
    $$
    \alpha = \arctan ( \F{\kappa}{\qk} ) \approx
    \alpha_0 + \cos^2 \! \alpha_0 \bittt
    \pmbf{(} \F{\kappa}{\qk} - \tan \alpha_0 \pmbf{)} \co
    \eqn\aads
    $$
we find
    $$
    R(d) \approx \qk \bitt \tan \pmbf{[} \alpha_0 - \beta_0 +
    \cos^2 \! \alpha_0 \bittt ( \F{\kappa}{\qk} - \tan \alpha_0 )
    \pmbf{]} \approx
    \cos^2 \! \alpha_0 \bittt ( \kappa - \qk \tan \alpha_0 \bitt ) \pe
    \eqn\aadt
    $$
Hence the choice,
    $$
     \kappa \sim ( \beta_0 \tan \beta_0 ) \bitt d^{-1} -
     \FF{1}{ \hskip -1.5pt \cos^2 \! \beta_0 } \bitt \fff{1}{L}
    \eqn\aadu
    $$
may lead to (\aadl). A possible regularized potential realizing (\aadu) is
    $$
    V_1(d) = \fff{\h^2}{2m} \[ \bit ( \beta_0^2 \tan^2 \!
    \beta_0 ) \bittt d^{-2} - \fff{2}{L} \bitt
    \FF{\beta_0 \biT \tan \beta_0}{\! \cos^2 \! \beta_0}
    \bittt d^{-1} \] \co
    \hskip 15pt V_2(d) = - \fff{\h^2}{2m} \bitt \beta_0^2 \bittt d^{-2}\ ,
    \eqn\aadv
    $$
which can be shown to give $R(d)$ satisfying (\aadl).

\case{(iii) case \mm{\alpha_0 = \pi/2}}

We still have \mm{\qk \sim \beta_0 \bit d^{-1}} but now \mmm{\kappa/\qk
\to \infty} so \mmm{ \alpha \approx \F{\pi}{2} - \F{\qk}{\kappa} \co } and
therefore
    $$
    R(d) \approx \qk \tan \bigg[ \f{\pi}{2} - \f{\qk}{\kappa}
    - ( \beta - \beta_0 ) - \beta_0 \bigg] \approx \qk
    \bigg[ - \f{\qk}{\kappa} - ( \beta - \beta_0 ) \bigg] \pe
    \eqn\aadw
    $$
The realization (\aadl) will be attained if, for example, we have
\mm{\kappa / \qk^2 \to \infty} and provide \mm{ - \f{1}{L} } through
\mm{\qk} by assuming \mm{\qk \sim \beta_0 \bit d^{-1} + \f{1}{L}
\f{1}{\beta_0} }.  This is the case with the regularization,
    $$
    V_1(d) = \fff{\h^2}{2m} \bittt c_1^2 \bitt d^{\bit 2 \qnu} \hskip 15pt
    ( \qnu < -2 ) \co \hskip 29pt V_2(d) = - \fff{\h^2}{2m}
    \( \beta_0^2 \bit d^{-2} + \fff{2}{L} \bitt d^{-1} \)\ .
    \eqn\aadx
    $$

To summarize, the regularization by means of the step-like potential
(\aaaa) leads generically to the standard wall \mm{L = 0}.  It can also
lead to nonstandard walls \mm{L \ne 0} but only as exceptional cases under
the fine-tuning (\aadn).  It is worth emphasizing that the crucial factor
in determining the limit of $R(d)$, {\it i.e.}, the boundary condition at
$x = 0$, is not the leading asymptotic behavior of \mm{V_1} and \mm{V_2}
in $d \to 0$ but always a subleading term.  A similar phenomenon has been
observed for the regularization of the Dirac delta point interactions in
three space dimensions [\AGHH].

The regularizations we used are based on a step-like potential.  Needless
to say, other types of potentials can also be used for realizing the
walls.  One can, for instance, look for a potential which leads to the
realization for any $L$ without involving the mass parameter $m$.  Such a
regularization may be more desirable than that we constructed --- where
the potentials turned out to be $m$-dependent --- for the reason that
potentials should be independent of the particle.  Nonetheless, our simple
regularization may well exhibit a universal feature of the realization of
the (standard and nonstandard) walls, as we can see, for example, the
bound state being accommodated in the negative middle part of the
step-like potential we used.

\sectit{4. Classical counterparts}

\secno=4 \meqno=1

Having seen that the quantum walls characterized by $L$ can be realized by
means of regularized potentials, we now turn to the question whether those
walls have classical counterparts or not.  We investigate this in the
phenomena of time delay discussed in section 2, by asking if there is a
classical system with some appropriate potential $V(x)$ which can account
for the same amounts of time delay as those observed under the walls. Note
that systems with the regularized potentials discussed above are not
applicable for this purpose, because in those systems the time a classical
particle spends in a potential (\aaaa) tends necessarily to zero as $d
\to 0$ (since, as \m{V_2 \to - \infty}, the distance run by the
particle becomes zero while its velocity becomes infinity).

To find a potential for the classical particle that reproduces the quantum
time delay, we shall first consider the walls with $L > 0$.  In this case
the time delay (\td) is negative, and if the classical picture is
available, the incident particle with velocity magnitude $ v = \frac{\hbar
k}{m}$ must return earlier by
    $$
    | \tau | = \f{2L}{v} \; \f{1}{ 1 + \( \f{mL}{\h} \, v \)^2 }
    \eqn\aabk
    $$
than we would expect when it collided with the wall at $x = 0$. Observe
that, for small $v$ the (minus) delay $\vert \tau \vert$ approaches
$\frac{2L}{v}$. This suggests that a slow particle sees the wall at
(around) $x = L$, not $x = 0$. Consequently, the reflecting potential
$V(x)$ is expected to begin to grow at $x = L$. For definiteness, let us
search for the potential in the qualitative form as shown in Figure~2.
(This fixes an arbitrariness in the choice of the potential. As we will
see, demanding a positive, monotonically decreasing potential determines
the potential uniquely.) Now, let us introduce
    $$
    \tilde{\tau} = \frac{2L}{v} + \tau
    = \sqrt{2m L^2 E} \Big/ \Big(\fff{\hbar^2}{2 m L^2} + E\Big) \co
    \eqn\aabl
    $$
(where $E = {1\over 2}mv^2$ is the incoming energy) which is the time
spent by the particle in the region left to the point \mm{x = L}. Our
problem is then an inverse problem: Determine a potential $V(x)$ from a
given $ \tilde{\tau}(E)$ as a function of $E$. This can be answered if we
follow the well-known argument of Landau and Lifshitz [\LL] used for the
problem of determining a well-shaped potential from the period time with
which a particle moves.

\topinsert
\epsfxsize 5.5cm
\ifx\omitpictures N  \centerline{\epsfbox {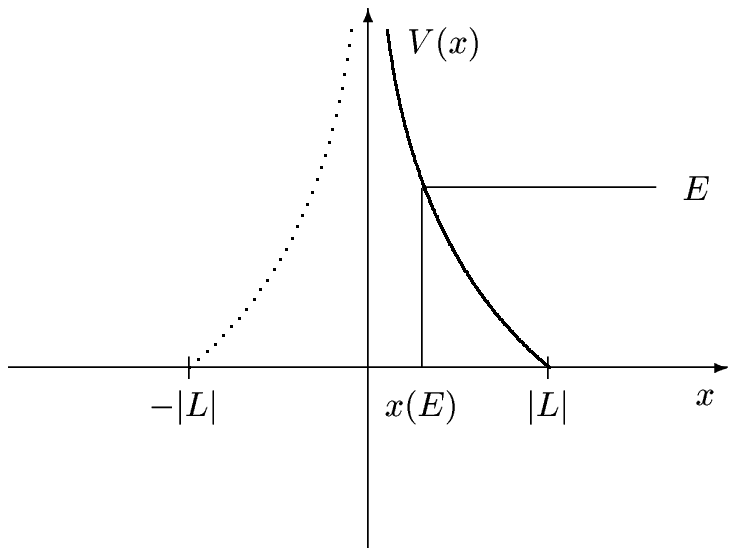}}  \fi
\vpt{3}
\abstract{{\bf Figure 2.}~The realizing potential (\AABR) is shown by the
solid line for $L > 0$.  For $L < 0$ the obtained potential becomes the
dotted line and is unphysical.}
\bigskip
\endinsert

We start by writing the relationship between the potential and
$\tilde{\tau}$ as
    $$
    \tilde{\tau} (E) = \sqrt{2m} \int_{ x(E) }^L
    \frac{\d x}{ \sqrt{ E - V(x) } } = \sqrt{2m} \int_0^E
    \( - \frac{ \d^{} x(V) }{\d V} \) \frac{\d V}{ \sqrt{E - V} } \pe
    \eqn\aabm
    $$
Dividing by $ \sqrt{W - E} $ with $W$ being an auxiliary parameter, and
integrating with respect to $E$ from 0 to $W$ leads to
    $$
    \int_0^W \f{ \tilde{\tau} (E) \,\, \d E }{ \s{W - E} }
    = \s{2m} \int_0^W \d V \( - \f{\d x}{\d V} \)
    \int_V^W \f{\d E}{ \s{ (W - E) (E - V) } } \pe
    \eqn\aabn
    $$
The inner integral (the one with respect to $E$) gives $\pi$, while on the
\lhs we can evaluate the integral explicitly [\cf (\aabl)]. {}From the
result,
    $$
    \pi \sqrt{2m} \, L \( 1 - 1 \Big/
    { \s{ 1 + \ffff{ 2 m L^2 }{\hbar^2} \, W } } \) =
    \pi \sqrt{2m} \bitt \[ L - x(W) \] \co
    \eqn\aabo
    $$
we obtain \mm{ x(W) = L \big[ 1 + \f{2mL^2}{\h^2} \bitt W
\big]^{-\f{1}{2}}\biTT, } inverting which yields\note{We remark that,
while this potential reproduces the time delay classically, it does not
reproduce the boundary condition (\aaag) and hence cannot serve as a
potential to realize the walls quantum mechanically.}
    $$
    V(x) = \fff{\h^2}{2 m L^2} \bittt \Big( \fff{L^2}{x^2} - 1 \Big) \pe
    \eqn\aabr
    $$

We can see that this wall-realizing potential sits on the positive half
line. This is unavoidable: Indeed, if a potential is identically zero on
the whole positive half line and is nonzero only on the negative half line
then the time delay is necessarily non-negative. The most we can reach is
that the penetration of the wall-realizing potential to the positive half
line is finite. (\aabr) presents such a solution. We will see that,
for \mm{L < 0}, we have to pay more.

For $L < 0$, the time delay is positive, \ie the quantum wave packet
returns later than expected:
    $$
    \tau = \f{2|L|}{v} \; \f{1}{ 1 + \big( \f{m|L|}{\h} \, v \big)^2 } =
    \s{2m} |L| \bitt \f{1}{ \s{E} \( 1 + \f{ 2 m L^2 }{\h^2} E \) } \pe
    \eqn\aack
    $$
This is the time delay we try to reproduce with the corresponding
classical particle as its classical time delay
    $$
    \taucl (E) = \s{2m} \int_{x(E)}^{x_0}
    \f{\d x}{ \s{E - V(x)} } - \f{2 \bit x_0}{ \s{2E/m} } \co
    \eqn\aaer
    $$
where \m{x_0} is the initial position of the particle. For small $v$,
(\aack) becomes $\f{2 |L|}{v}$, which suggests that a slow particle enters
the $x < 0$ region and sees the wall near \mm{x = - |L| \pe } For this,
the realizing potential $V(x)$ is expected to start to increase at $x = -
|L|$, and to keep increasing for smaller $x$. However, if one repeats the
same argument used for the $L > 0$ case, one ends up with (\aabr) again,
with now the left branch of this function (see Figure~2).  The obvious
problem with this branch, {\it i.e.}, it increases for $x$ to the right of
$-|L|$ and is unphysical, may be understood intuitively as follows.  For
high energies \m{E}, the particle is expected to move approximately
freely, and since the particle travels at least until \mm{x = - |L|}, the
\mm{ E \to \infty } asymptotics of the time delay would be at least \mm{
\f{2 |L|}{v} }. However, the time delay we have to reproduce has only a
\m{v^{-3}} asymptotic behavior.  This means that the coefficient of the
\m{v^{-1}} term must vanish for \mm{ E \to \infty }, imlplying that in the
limit the particle reaches only until \mm{x = 0}. 

The situation cannot be helped with any additional potential in \mm{ - |L|
< x < 0 } or in \mm{0 < x \co } nor by any other modification. Actually,
it can be proven that no classically acceptable reflecting potential can
fulfil the requirement that the time delay (\aack) be reproduced exactly
for all \mm{x_0 >\xl}, that is, for all initial positions of the incoming
particle above a finite, possibly positive threshold position \mm{\xl}. 
To see this, let us consider an arbitrary piecewise differentiable
potential, even possibly diverging at the discontinuity points. Then the
classical force \mm{ - V'(x) } exists everywhere except for finitely many
points, while at a discontinuity point an incoming classical trajectory
can be continued with the outgoing trajectory that has the same energy
\mm{E} as the incoming one. The potential is further required to act as a
completely reflecting wall, that is, for every positive energy \m{E},
there has to be a turning point \mm{ x(E) } (like in Fig.~2). Note that
then the function \mm{ x(E) } is necessarily nonincreasing, and its
inverse is \m{V(x)} locally, \ie it reproduces at least parts of the
function \m{V(x)}. 

First let us discuss the case when \m{V} is differentiable (and hence
continuous) everywhere. The \m{x_0}-independence of the time delay \mm{
\taucl(E) } [\cf (\aaer)] implies
    $$
    \f{\d}{\d x_0} \bitt \taucl (E) =
    \s{2m} \[ \ff{1}{ \s{E - V(x_0)} } - \f{1}{ \s{E} } \] = 0
    \eqn\aaem
    $$
and thus that \mm{V = 0} above \mm{\xl}. Let \m{\xo} denote the lowest
\m{x} above which the potential is nonpositive. Naturally, one has \mmm{
\xo \le \xl } and can write \mmm{ \xo = \sup \bitt \{ x \bitt | \bitt V(x)
> 0 \} \co } from which one finds \mmm{ \xo = \lim_{E \downto 0} \bitt
x(E)}, that is, \m{\xo} is the \quotess{turning point for zero energy}. 

If there exists an energy \m{\Eo} with a turning point on the negative
half line, \mm{ x(\Eo) < 0 \co } then for larger energies \m{E} the time
delay is at least
    $$
    \s{2m} \int_{ x(\Eo) }^{x_0}
    \f{\d x}{ \s{E - V} } - \s{2m} \bitt \f{\hpt{4} x_0}{ \s{E} }
    \eqn\aaen
    $$
which is obtained by omitting the time of travelling through the interval
\mm{ [ \bit x(E), \bittt x( \Eo ) \bit ] }. Since \m{V} is continuous on
the interval \mm{ [ \bit x(\Eo), \bitt x_0 \bit ] }, it is bounded and
hence the high-energy asymptotics of (\aaen) is
    $$
    \s{2m} \bittt \f{ x_0 - x(\Eo) }{ \s{E} } -
    \s{2m} \bitt \f{ \hpt{4} x_0 }{ \s{E} } =
    \s{2m} \bittt \f{ | x(\Eo) | }{ \s{E} }
    \sim \f{ \hpt{1} 1 }{ \s{E} } \pe
    \eqn\aaeo
    $$
This is in contradiction with the asymptotics \mm{ E^{- 3/2} } of the
demanded time delay (\aack). Consequently, all turning points have to be
on the non-negative half line,
    $$
    x(E) \ge \lim_{E' \to \infty} x(E') =: \xinf \ge 0 \pe
    \eqn\aaep
    $$

Next we prove that in \m{( \bit \xinf, \bittt \xo \bit ] \bitttt} the
potential \m{V \biT} decreases strictly. Namely, if we assume the contrary
then there will be at least one point \m{\xone} in this interval that is
not a turning point (see Fig.~3a). Within \m{ [ \bit \xone, \bitt \xo \bit
] }, let \m{\xtwo} denote the turning point with the highest energy
\m{\Etwo}. Then, in the function \m{\taucl (E) } there will be a
discontinuity at \mm{E = \Etwo}: 
    $$
    \ff{1}{ \s{2m} }
    \[ \lim_{E \downto E_2} \taucl (E) - \lim_{E \upto E_2} \taucl (E) \]
    = \lim_{E \downto E_2} \int_{x(E)}^{x_0} \ff{\d x}{ \s{E - V} }
    - \lim_{E \upto   E_2} \int_{x(E)}^{x_0} \ff{\d x}{ \s{E - V} }
    $$
    $$
    = \lim_{E \downto E_2} \[ \int_{x(E)}^{\xtwo} \ff{\d x}{ \s{E - V} }
    + \int_{\xtwo}^{x_0} \ff{\d x}{ \s{E - V} } \]
    - \lim_{E \upto E_2} \int_{x(E)}^{x_0} \ff{\d x}{ \s{E - V} }
    \eqn\aaeq
    $$
    $$
    \hpt{15}
    = \lim_{E \downto E_2} \int_{x(E)}^{\xtwo} \ff{\d x}{ \s{E - V} }
    > \lim_{E \downto E_2} \int_{\xone}^{\xtwo} \ff{\d x}{ \s{E - V} }
    = \int_{\xone}^{\xtwo} \ff{\d x}{ \s{E_2 - V} } > 0 \pe
    $$
However, the required quantum time delay, (\aack), is a continuous
function everywhere.  This result tells us that on the region \mmm{ ( \bit
\xinf, \bittt \xo \bit ] } \mm{x(E)} is the inverse of \m{V(x)} and is
differentiable. We have also obtained the qualitative behavior of the
candidate potential function (see Fig.~3b): Coming from the right, it is
zero above \m{\xl}, nonpositive in \mm{ \xo < x < \xl \bit , } and is
positive and increasing in \mm{ \xinf < x \le \xo \bit , } diverging to
\mm{+\infty} at \mm{\xinf \pe}

\topinsert
\epsfxsize 12.7cm
\ifx\omitpictures N  \centerline{\epsfbox{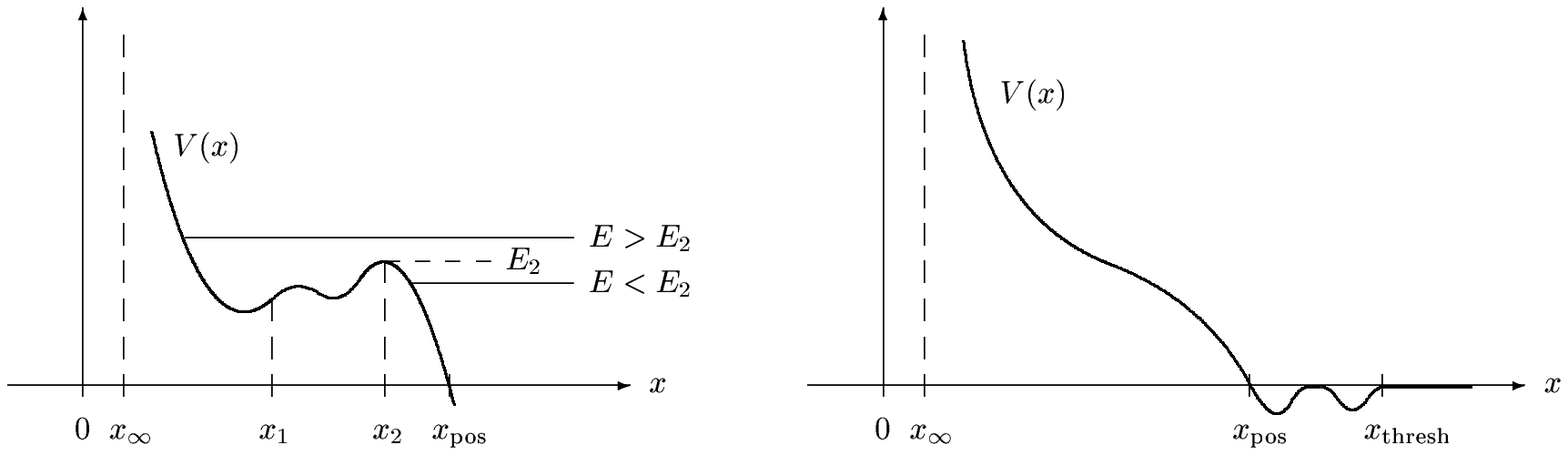}}  \fi
\vpt{5}
\centerline{\abstract{a) \hskip 50ex b)}}
\vskip 2ex
\abstract{{\bf Figure 3.}
\hpt{4} a) A nondecreasing part in the potential in \mmm{ ( \bit \xinf,
\bittt \xo \bit ] } causes a discontinuity in the time delay.
\hpt{7} b) The obtained qualitative shape of the potential.}
\bigskip
\endinsert

Now we are ready to investigate the requirement \mmmm{ \f{1}{ \s{2m} }
\bitt \tau (E) = \f{1}{ \s{2m} } \bitt \taucl (E) \, : }
    $$
    \f{ |L| }{ \s{E} \( 1 + \f{ 2 m L^2 }{\h^2} E \) } =
    \int_{x(E)}^{\xo} \f{\d x}{ \s{E - V} } +
    \int_{\xo }^{x_0} \f{\d x}{ \s{E - V} } -
    \f{\hpt{4} x_0}{ \s{E} } \pe
    \eqn\aaet
    $$
Observe that the second integral is bounded from above by \mmm{ \f{ x_0 -
\xo }{ \s{E} } \co } since the potential is nonpositive on that interval.
Employing again the \quotes{Landau trick} to the first integral (\ie
changing the variable from \m{x} to \m{V}, dividing by \mm{ \sqrt{W - E}
\co } and integrating between \m{0} and \m{W}), we find
    $$
    \pi \bi |L| \Big/ { \s{ 1 + \ffff{ 2 m L^2 }{\h^2} W } } \le
    - \pi \bit x(W) \co
    \eqn\aaeu
    $$
or
    $$
    x(W) \le
    - { |L| } \Big/ { \s{ 1 + \ffff{ 2 m L^2 }{\h^2} W } } < 0 \pe
    \eqn\aaeu
    $$
This, however, contradicts our previous result that all turning points have
to be on the non-negative half line, showing that the requirement (\aaet)
cannot be fulfilled.

We can show that the preceding argument remains valid even if we allow
discontinuity points in the potential --- only slight modifications are
necessary. The \m{x_0}-independence of the time delay implies \mm{V = 0}
at all continuity points, and hence everywhere, above \m{\xl}. \m{\xo} is
introduced in the same way and with the same properties as before. (\aaep)
also remains valid: When assuming \mm{ x(\Eo) < 0 \co} the possible
discontinuity points falling between \m{x(\Eo)} and \m{x_0} can be covered
by intervals of a total length less than, say, \mm{ \f{1}{2} | x(\Eo) | }.
We omit even these covering intervals from the time delay, and on the
remaining intervals the potential is continuous and has overall upper and
lower bounds. Consequently, the high-energy asymptotics of the time delay
is still at least \mm{ \sim 1 \big/ \s{E} \pe }

The proof of the strict decreasing of \m{V} in \mm{ ( \bit \xinf, \bittt
\xo \bit ] } holds, too. This also rules out discontinuity points \mm{\xd}
in \mm{ ( \bit \xinf, \bittt \xo \bit ] } with \mmmm{ V (\xd - 0) < V (\xd
+ 0) \pe } Others are allowed but do not cause any trouble in the behavior
of \m{x(E)} because, for energies \mm{ E \in [ \bit V (\xd + 0), \bitt V
(\xd - 0) \bitt ] }, we then have \mm{ x(E) = \xd = \const } and \mmm{
\f{\d}{\d E} \bitt x(E) = 0 \pe } The transformation of the integration
variable in the first integral in (\aaet) remains applicable, while the
second integral can also be estimated as before, in spite of any
discontinuity points in \m{ ( \bit \xo, \bittt \xl \bit ] \pe } Therefore,
we reach the same contradictory result (\aaeu) again. 

Hence, interestingly enough, the walls with negative $L$ do not admit a
classical counterpart, {\it i.e.}, they are genuinely quantum. 
Incidentally, we mention that if we demand only that the quantum time
delay be reproduced in the \mmm{x_0 \to \infty} limit of \mm{ \taucl (E)
\co } then the required realization can be achieved (see the Appendix).

\sectit{5. WKB-exactness}

\secno=5 \meqno=1

The fact that for walls with $L = 0$ and $L = \infty$ the transition
kernel is almost WKB-exact alludes us to examine whether this implies a
complete exactness or not, and if so, whether such a feature persists to
nonstandard walls as well. More precisely, we wish to see if the sum of
amplitudes along the classical two paths, the direct world line from $(x,
t) = (a, 0)$ to $(b, T)$ and the bouncing path which hits the wall $x = 0$
before arriving at $(b,T)$, give the exact result (see Figure~4a). The
question, therefore, is if the kernels (\aabd), (\aabe) and (\aabf) can be
rewritten in the form of a sum of the corresponding two terms as
    $$
    K(b, T; a, 0) = \s{ \ff{m}{2 \pi i \h T} } \bittt
    e^{ \f{i m}{2 \h T} (b - a)^2 } + \s{ \ff{1}{2 \pi i \h}
    \ff{\p^2 \Sb}{\p a \bitt \p b} } \bittt
    e^{ \f{i}{\h} \Sb (b, T; a, 0) } \co
    \eqn\aaco
    $$
where \mm{ \Sb (b, T; a, 0) } is the classical action for the bounce path,
and the factor before the second exponential term comprises the van Vleck
determinant and the Maslov phase factor corresponding to the turning point
(see [\Schulman] for the details). In the spirit of the preceding
sections, here again the wall is considered not necessarily to be simply
the infinitely high vertical potential wall at the origin but to be
realized by some sequence of more general reflecting potentials. What we
require is that the potential sequence must converge uniformly to zero for
all \mm{x \ge \xk} with some \mm{\xk} which may be positive, and that, for
any \mm{ a, b > \xk , } the bounce path tends to the standard bounce world
line depicted on Fig.~4a, at least on the spacetime region \mm{x > \xk}. 
Otherwise we let the reflecting potential sequence be arbitrary to the
left of \m{\xk} and, therefore, at the limit of the sequence, the
resultant action \mm{ \Sb } can differ from the action \mm{ \Sb^{(0)} =
\f{ m (a + b)^2 }{2 T} } that corresponds to the simplest case of the
infinitely high vertical potential wall with no extra action contribution
caused by the wall. 

\topinsert
\epsfxsize 11.4cm
\ifx\omitpictures N  \centerline{\epsfbox {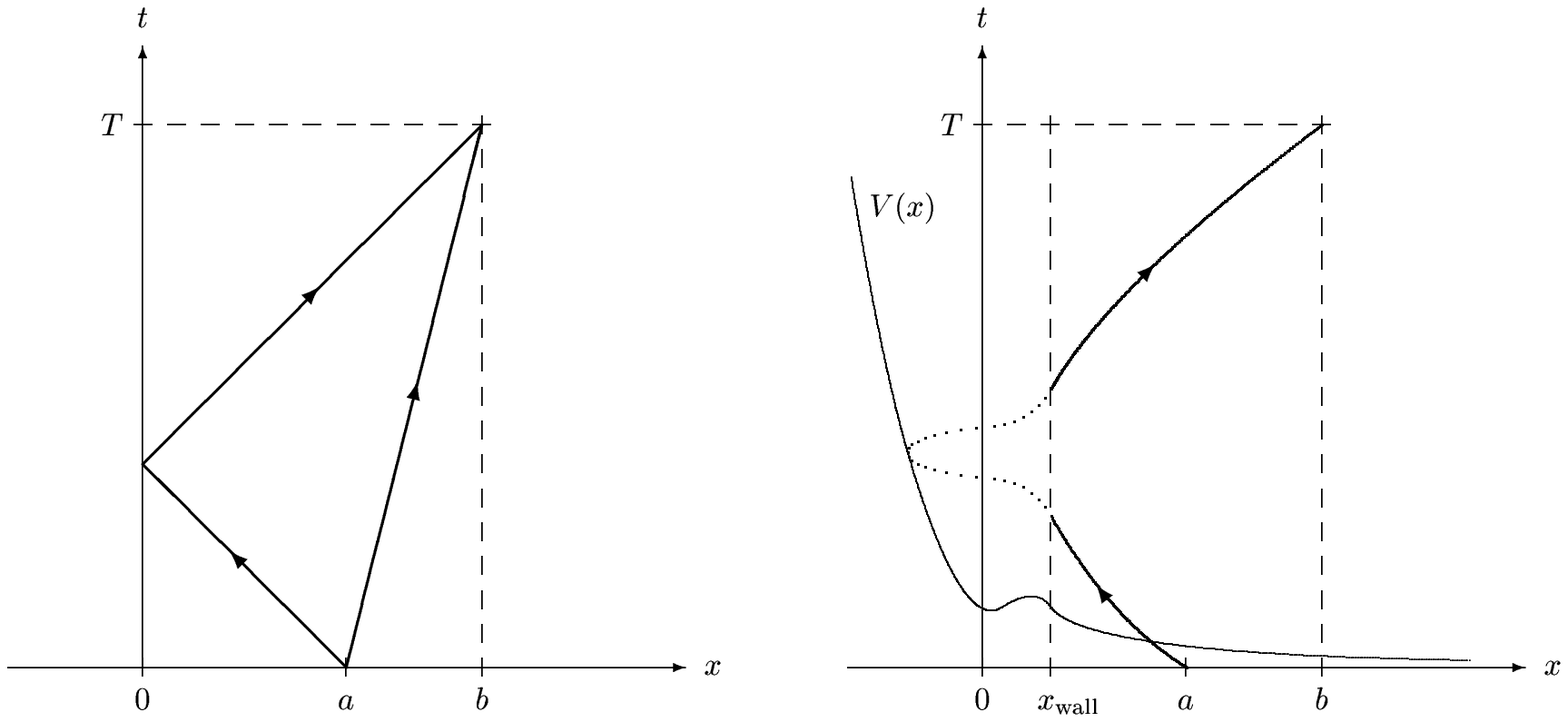}}  \fi
\abstract{ \phantom{.} \hpt{67} a) \hpt{197} b)}
\vskip 2ex
\abstract{{\bf Figure 4.} $\,$ a) The direct and the bounce paths.
$\,$ b) The bounce path under a wall-realizing potential.}
\bigskip
\endinsert

Even these very mild assumptions allow us to observe some important,
generally valid, properties. The first one is that, although the direct
path is also influenced by a nonvanishing potential, its WKB contribution
\mmm{ \s{ \f{i}{2 \pi \h} \f{\p^2 \Sd}{\p a \bitt \p b} } \bittt e^{
\f{i}{\h} \Sd } } will still reduce to the first term of (\aaco).  Indeed,
since in the limit we have $V \to 0$ and hence the velocity of the
particle tends uniformly to \mm{ \f{b - a}{T} }, we trivially find \mmm{ E
\to \Ed^{(0)} = \f{m}{2} \f{ (b - a)^2 }{ T^2 } } and \mmm{ \Sd^{} \to
\Sd^{(0)} = \f{m}{2} \f{ (b - a)^2 }{T} }.  The nontrivial question that
remains to be shown concerns with the property of the derivative, \mmm{
\f{ \p^2 \Sd^{} }{\p a \bitt \p b} \to \f{\p^2 \Sd^{(0)} }{\p a \bitt \p
b} }, but this can be seen by writing the action as
    $$
    \Sd = \int_0^T \d t \, (E - 2V) = - T E + 2 \int_0^T \d t \,
    (E - V) = - T E + \s{2m} \int_a^b \d x \bit \s{E - V}
    \eqn\aacl
    $$
which is valid for \mm{ a, b > \xk , } and evaluating
    $$
    \f{ \p^2 \Sd }{ \p a \bitt \p b } =
    - \f{ \s{m/2} }{ \s{ E - V(a) } } \f{ \p E }{ \p b } =
    - \s{2m} \[ \s{ E - V(a) } \s{ E - V(b) }
    \int_a^b \f{ \d x }{ \s{ E - V }^3 } \]^{-1} \biTTTT .
    \eqn\aaex
    $$
Here, the energy \m{E} of the direct path is determined by the condition
    $$
    \s{ \fff{m}{2} } \int_a^b \f{ \d x }{ \s{E - V} } = T
    \eqn\aaev
    $$
which is used to evaluate \mmm{ \f{ \p^2 \Sd }{\p a \bitt \p b} } in
(\aaex).  Plugging the limiting values for the energies and the potential
in (\aaex), we find the required property. 

Second, we make the observation that the energy of the bounce path
converges to \mm{ \Eb^{(0)} = \f{m}{2} \f{ (a + b)^2 }{T^2} }.  This
follows from our requirement that the bounce path must tend to the
standard bounce world line outside \mm{\xk} because then the velocity of
the particle tends uniformly to \mm{ \f{a + b}{T} } under the vanishing
potential.  In addition, we find that, although \mmm{ \Delta \Sb = \Sb -
\Sb^{(0)} } does not necessarily tend to zero, in the limit it becomes
independent of \m{a} and \m{b}. This can be seen as follows: 
    $$
    \Sb = - T E + \s{2m} \int_{x(E)}^a \d x \bit \s{E - V} +
    \s{2m} \int_{x(E)}^b \d x \bit \s{E - V}
    \eqn\aaew
    $$
and
    $$
    \F{ \p \Sb }{ \p a } = \s{2m} \bit \s{ E - V(a) } \co
    \eqn\aaey
    $$
where now the energy of the bounce path is determined by
    $$
    \s{ \fff{m}{2} } \int_{x(E)}^a \f{ \d x }{ \s{E - V} } +
    \s{ \fff{m}{2} } \int_{x(E)}^b \f{ \d x }{ \s{E - V} } = T
    \eqn\aaez
    $$
[again, (\aaez) is used also for the result (\aaey)]. Since now \mmm{ E
\to \Eb^{(0)} \co } it follows that
    $$
    \F{ \p \Sb^{} }{ \p a } \bitttt \to \bitttt m \FGF{a + b}{T} =
    \F{ \p \Sb^{(0)} }{ \p a }
    \eqn\aafg
    $$
so \mm{ \F{ \p \Delta \Sb^{} }{ \p a } \to 0 \pe } The \m{b}-independence
of \mm{ \Delta \Sb } is proven analogously.

Third, if we restrict ourselves to the potential sequences of the type
(\aaaa) then (\aaew) and (\aaez) read
    $$
    \Sb = - T E + \s{2m} \bitt (a + b) \s{E} + 2 \s{2m} \hpt{2}
    d \bit \s{E + |V_2|}
    \eqn\aafa
    $$
and
    $$
    \s{ \F{m}{2} } \hpt{5} \f{ a + b }{ \s{E} } +
    \s{ \F{m}{2} } \hpt{3} \f{d}{ \s{E + |V_2| } } = T \pe
    \eqn\aafb
    $$
{}From (\aafa) we have that
    $$
    \lim_{d \to 0} \Delta \Sb =
    \lim_{d \to 0} \( 2 \s{2m} \hpt{2}d \s{ |V_2| } \) \pe
    \eqn\aafc
    $$
In parallel, \mmm{ \FFF{\p^2 \Sb}{ \p a \bitt \p b } } can be computed by
differentiating (\aaey), and using \mm{ \F{\p E}{\p b} \co } the latter
obtained by expressing \mm{b = b(E)} from (\aafb) and applying \mmm{ \F{\p
E}{\p b} = 1 / [ \F{\p b}{\p E} ] \pe } Taking the limit of the result
gives \mm{ m / T } so we find that, in the limit, the square root factors
in the two terms of (\aaco) equal each other for these step-like potential
sequences.

By virtue of these properties, we are able to discuss the question of
complete WKB-exactness. In the cases $L = 0$ and $L = \infty$, it is
possible to reproduce the required action contribution \mmm{ \Delta \Sb =
\pi \h } and \mmm{ \Delta \Sb = 0 \co } respectively, for example with the
step-like potential sequences (\aaaa). In fact, choosing for \m{L= 0}
    $$
    V_1(d) = \const d^{-1} \co \qquad
    V_2(d) = - \fff{\h^2}{2m} \( \fff{\pi}{2} \)^2 d^{-2}
    \eqn\aabj
    $$
(a potential sequence with \mm{\alpha_0 = 0} and \mm{\beta_0 = \pi/2}),
and for $L = \infty$
    $$
    V_1 (d) = \fff{\h^2}{2m} \bitt \qc^2 \bitt d^{-1} \co \hpt{25}
    V_2 (d) = - \fff{\h^2}{2m} \bitt \qc \bittt d^{ - \f{3}{2} } \co
    \eqn\linfaadr
    $$
which is the case \mmm{\nu = - 1/2} of (\aadr), provides just these needed
action contributions [\cf (\aafc)]. Note that these potential sequences
are, at the same time, correct realizations of the quantum boundary
condition with \mm{L = 0 \co } respectively \mm{ L = \infty \co} as well.
Nevertheless, they are not unique even among the step-like realizations
with these properties, and presumably other potential shapes can also
serve as examples for even both the complete WKB-exactness and realizing
the quantum boundary condition.

On the other side, for the other walls \mm{ L \ne 0, \infty \co } one can
prove that no potential sequence can account for the kernels (\aabe)
and (\aabf) irrespective of whether the potential sequence reproduces the
correct quantum boundary condition or not. To see this, let us write these
kernels in the form
    $$
    \s{ \fff{m}{2 \pi i \h T} } \[ e^{ \f{i m}{2 \h T} (b - a)^2 } +
    A_L (a, b, T) \bitt e^{ \f{i}{\h} \Sb^{(0)} } \] \pe
    \eqn\aafe
    $$
If the complete WKB-exactness holds then \mmm{\arg A_L (a, b, T) } should
correspond to the limit of \mmm{ \Delta \Sb / \h \co } which we know is
unavoidably independent of \m{a} and \m{b}. However, actually \mmm{ \arg
A_L (a, b, T) } does depend on \m{a} and \m{b}, as can be checked simply
for example, on the large-\m{T} asymptotics of \mm{ A_L (a, b, T) \co }
    $$
    A_L (a, b, T) \approx \mdef{ - \s{ \f{m}{2 \pi i \h T} }_{\ph{|}}
    \bitt e^{ - \f{2imL}{\h T} ( a + b - L ) } \co & \hpt{15} L < 0
    \co \cr {\f{2}{L}}^{\ph{|}} e^{ - \f{a + b}{L} } \bitt
    e^{ - \f{im}{2 \h T} \[ (a + b)^2 - \( \f{\h T}{mL} \)^2 \] }
    \bit , & \hpt{15} L > 0 \co }
    \eqn\aaff
    $$
as one finds from (\aabe) and (\aabf).

We thus learn that the quantum walls with $L = 0$ and $L = \infty$, which
correspond to the Dirichlet $\psi(0) = 0$ and the Neumann $\psi'(0) = 0$
boundary condition, respectively, are distinguished in the $U(1)$ family
of walls with respect to the WKB-exactness. These two cases are
distinguished also by their scale invariance which arises due to the
absence of the scale parameter. The relationship between the two, the
WKB-exactness and scale invariance, is however unclear.

\bigskip
\noindent
{\bf Acknowledgements:}
This work has been supported in part by the Grant-in-Aid for Scientific
Research (C)  (Nos.~11640396 and 13640413) by the Japanese Ministry of
Education, Culture, Sports, Science and Technology.

\vfill\eject
\sectit{Appendix: Weak classical realization of the time delay for
\m{ \pmbf{L < 0} } }

\secno=0 \appno=1 \meqno=1

Here, we outline how a weaker classical realization of the quantum time
delay, namely, as the \mmm{x_0 \to \infty} limit of the classical time
delay \mm{ \taucl (E) \co } can be determined for the walls \m{L < 0}. Let
us assume that we have a strictly decreasing positive potential such that,
for a fixed finite \m{x_0} and all energies \m{E} above \m{V(x_0)}, \mmmm{
\taucl (E) = \tau(E) \pe } We use the \quotes{Landau trick} again,
dividing this equation by \mm{ \sqrt{W - E} \co } integrating now between
\m{ V(x_0) } and \m{W}, and evaluating the left hand side by changing the
variable to \m{V}. {}From the result we can express \mm{x(W) \biTT } to
find 
    $$
    x(W) = \ff{x_0}{\pi} \arccos \[ 1 - \ff{ 2 V(x_0) }{W} \] -
    \ff{ 2 |L| / \pi}{ \s{1 + \f{2 m L^2}{\h^2} W} } \arccos
    \s{ \ff{ 1 + \f{2 m L^2}{\h^2} W }{ 1 + \f{2 m L^2}{\h^2} V(x_0) }
    \ff{ V(x_0) }{W} } \pe
    \eqn\aaec
    $$
Now we perform the limit \mm{ x_0 \to \infty \co } with a fixed \m{W}. The
second term on the \rhs of (\aaec) remains finite no matter how
\m{V(x_0)} changes correspondingly. Consequently, to have a finite
\mm{x(W)} in the limit, \mmmm{ \arccos \[ 1 - \f{ 2 V(x_0) }{W} \] } has
to tend to zero.  This means \mm{ V(x_0) \to 0 \co } and from \mmmm{ \cos
\eps \approx 1 - \f{\eps^2}{2} \quad ( \eps \approx 0 ) } we have the
asymptotics
\mmmm{ \arccos \[ 1 - \f{ 2 V(x_0) }{W} \] \approx 2 \s{ \f{V(x_0)}{W} } }
so to reach a finite limit of (\aaec) \mmm{ x_0 \s{ V(x_0) } } has to
converge to a constant. Introducing
    $$
    c := \lim_{x_0 \to \infty} \bi \fff{ 2 \s{2m} }{\pi \h} \bitttt
    x_0 \bi \s{ V(x_0) } \ ,
    \eqn\aaee
    $$
which will be a free parameter in the realizing potential, the limit of
(\aaec) is
    $$
    x(W) = \fff{\h}{ \s{2m} } \( c \big/ \s{W} \bittt - \bittt
    1 \Big/ \s{ \ffff{\h^2}{2 m L^2} + W } \) \pe
    \eqn\aaef
    $$
One can check that the inverse of this \mm{x(W)} is really a strictly
decreasing potential tending to zero if \mm{c \ge 1 \co} and that the time
delay corresponding to it is
    $$
    \taucl (E) = \h \biT \[ \bitt c \bittt
    \ff{ \s{ 1 - \f{ V(x_0) }{E} } - 1 }{ \s{ V(x_0) E } } +
    \ff{1}{ \s{ \f{\h^2}{2 m L^2} + V(x_0) } } \( \ff{1}{ \s{E} }
    - \ff{ \s{ E - V(x_0) } }{ \f{\h^2}{2 m L^2} + E } \) \] \biTT ,
    \eqn\aaeg
    $$
whose \mm{x_0 \to \infty} limit is really the desired quantum time delay
(\aack) (independently of \m{c}).
The potential itself is obtained by solving the biquadratic equation that
follows from (\aaef), and reads, for example, for \mm{ c = 1 ,}
    $$
    V(x) = \fff{2 \h^2}{m L^2} \( \fff{x}{|L|} \)^{ - \f{2}{3} }
    \[ \( \fff{x}{|L|} \)^{ \f{2}{3} } + \qn^{-1} +
    2 \s{\qn - \qn^4} \bitt \]^{-2}
    \eqn\aaek
    $$
with
    $$
    \qn = \fff{1}{ \s{2} } \[ \( \s{ 1 + \fff{1}{27} \(
    \fff{x}{|L|} \)^4 } + 1 \)^{ \f{1}{3} } - \( \s{ 1 + \fff{1}{27}
    \( \fff{x}{|L|} \)^4 } - 1 \)^{ \f{1}{3} } \]^{ \f{1}{2} } \pe
    \eqn\aael
    $$


\baselineskip= 15.5pt plus 1pt minus 1pt
\parskip=5pt plus 1pt minus 1pt
\tolerance 8000
\vfill\eject\immediate\closeout\reffile
\centerline{{\bf References}}\bigskip\frenchspacing%
\input refs.tmp\vfill\eject\nonfrenchspacing

\bye